\documentclass[11pt,a4paper]{article}
\pdfoutput=1

\usepackage{amsmath}
\usepackage{amsfonts}
\usepackage{amsthm}
\usepackage{amssymb}
\usepackage{amscd}
\usepackage[british]{babel}
\usepackage{graphicx}
\usepackage{psfrag}
\usepackage{epsfig}
\usepackage{rotating}
\usepackage{times}
\usepackage{color}
\usepackage{ulem}

\theoremstyle{plain}

\newtheorem{remark}{Remark}[section]

\newcommand{\boxend}{\flushright{$\Box$}}

\newcommand{\R}{{\mathbb R}}               % for real numbers
\renewcommand{\tilde}{\widetilde}

\begin{document}

%\title{Matter Bounce Scenario: current status and new challenges}

\title{An Extended Matter Bounce Scenario: current status and challenges}

\author{Jaume de Haro$^{a,}$\footnote{E-mail: jaime.haro@upc.edu} and Yi-Fu Cai$^{b,}$\footnote{E-mail:yifucai@physics.mcgill.ca}
}

\maketitle

\begin{center}
{\small
$^a$Departament de Matem\`atica Aplicada I, Universitat
Polit\`ecnica de Catalunya \\ Diagonal 647, 08028 Barcelona, Spain \\
$^b$Department of Physics, McGill University, Montr\'eal, QC, H3A 2T8, Canada
}
\end{center}

%&&&&&&&&&&&&&&&&&&&&&&&&&&&&&&&&&&&&&&&&&&&&&&&&6

\thispagestyle{empty}

\begin{abstract}

As an alternative to the paradigm of slow roll inflation, we propose an
extended scenario of the matter bounce cosmology in which the Universe
has experienced a quasi-matter contracting phase with a variable background
equation of state parameter. This extended matter bounce scenario can be
realized by considering a single scalar field evolving along an approximately
exponential potential. Our result reveals that the rolling of the scalar field in
general leads to a running behavior on the spectral index of primordial cosmological
perturbations and a negative running can be realized in this model.
We constrain the corresponding parameter space by using the newly released
Planck data. To apply this scenario, we revisit bouncing cosmologies within
the context of modified gravity theories, in particular, the holonomy corrected
loop quantum cosmology and teleparallel $F(T)$ gravity. A gravitational process
of reheating is presented in such a matter bounce scenario to demonstrate
the condition of satisfying current observations. We also comment on several
unresolved issues that often appear in matter bounce models.

\end{abstract}

{\bf Pacs numbers:} 04.50.Kd, 98.80.Bp, 98.80.Jk

\maketitle

\section{Introduction}

The Matter Bounce Scenario (MBS) (see \cite{Cai:2014bea, Brandenberger:2009jq, Novello:2008ra, Lehners:2008vx, Battefeld:2014uga} for recent reviews on a variety of 
bouncing cosmologies) suggests that the Big Bang singularity is replaced by a nonsingular bounce. It is essentially characterized by the Universe being nearly matter 
dominated at very early times in the contracting phase (to obtain an approximately scale invariant power spectrum) and evolving towards a bounce where all the parts of the Universe 
become in causal contact \cite{Amoros:2013nxa}, solving the horizon problem, to enter afterwards into a regular expanding phase, in which it matches the behavior of the standard 
hot big bang cosmology. As an alternative to the slow roll inflationary paradigm, the MBS can be free from the problem of the initial singularity \cite{Borde:1993xh} that inflation 
models suffer from. Also, the potential for the scalar field in MBS does not need to be extremely flat as what is required in inflationary cosmology 
\cite{Adams:1990pn}.

In order to obtain a viable MBS model that can compete with the inflationary paradigm, it is expected that the underlying model can satisfy a variety of theoretical
and observational constraints. Moreover, there exist several conceptual issues that are not clear in the frame of the MBS. We list these points in the following and discuss 
the conditions of model building in the MBS in the present work.

First, today's cosmological measurements, such as the Planck data released in 2013 (Planck2013) \cite{Ade:2013zuv,Ade:2013uln} as well as in 2015 
(Planck2015) \cite{Planck:2015xua, Planck:2015xub}, have precisely determined the amplitude of the power spectrum for primordial curvature perturbations to
be ${\mathcal P}_{\xi}\cong 2.2 \times 10^{-9}$. This amplitude, in bounce models, is often associated with the energy scale of the bounce as well 
as the process of primordial perturbations evolving through the nonsingular bouncing phase \cite{Cai:2014xxa}. Accordingly, the theoretical result of the 
power spectrum of primordial curvature perturbation calculated from any bounce model has to match with the observational data.

Second, according to the Planck2013 data, at $1\sigma$ confidence level (C.L.) the spectral index for curvature perturbation and its running, namely, $n_s$ and $\alpha_s$, are 
constrained to be $0.9603\pm 0.0073$ and $-0.0134\pm 0.009$, respectively \cite{Ade:2013zuv,Ade:2013uln}. These results are obtained upon the standard Lambda Cold Dark 
Matter ($\Lambda$CDM) model with a parameterized power spectrum and thus, are general to any inflation and alternative models. It is well-known that the ways to obtain 
a nearly scale invariant power spectrum of primordial perturbations are either a quasi de Sitter phase in the expanding phase or a nearly matter dominated phase in 
the contracting phase \cite{Wands:1998yp}. However, in an exact MBS with the background equation of state being $w=0$, one finds $n_s=1$ for the single
field model and therefore, conflicts with observations (which is similar to the fact that an exact de Sitter inflation was ruled out by the data). In order to improve 
the scenario to correctly 
explain observations, as in inflationary cosmology where a dynamically slow roll period was introduced, one expects a quasi matter dominated phase in bouncing 
cosmology with the background equation of state deviating from zero by the condition $\left| w\equiv \frac{P}{\rho}\right|\ll 1$, where $P$ and $\rho$ the pressure 
and the energy density of the Universe, must be considered at early times in the contracting phase.

Third, it is important to take into account the observational constraint of primordial non-gaussianity upon early Universe models. So far there is 
no evidence pointing to the existence of these nonlinear fluctuations \cite{Ade:2013ydc, Ade:2015ava}. As a result a large number of inflation models 
are ruled out by this observational fact. Thus, the no-detection of primordial non-gaussianity is expected to tightly constrain bounce models. For instance, it was studied 
in detail in \cite{Cai:2009fn} that primordial non-gaussianity generated in the simplest version of MBS is not sizable enough to be 
probed in cosmological observations. However, for the bounce models achieved by including a spatial curvature term, the amplitude of the nonlinear 
fluctuations could be dangerous to explain latest observations \cite{Gao:2014hea}. In general, the signature of the primordial non-gaussianity via the calculation of the 3-point 
function remains unclear in a generic picture of MBS since the calculation relies on the detailed process for primordial fluctuations evolving through 
the bouncing phase which contains uncertainties depending on different mechanisms of generating a nonsingular bounce.

Fourth, the latest CMB experiments including the BICEP2/Keck Array and Planck data have constrained the tensor-to-scalar ratio to be $r\leq 0.12$ with a pivot 
scale of $0.05$ ${\rm Mpc}^{-1}$ at $2\sigma$ C.L. \cite{Ade:2015tva}. When applied to inflation models, due to a slow roll {\it consistency} relation 
${r}=16\bar{\epsilon}$ \cite{Sasaki:1995aw} (where $\bar{\epsilon}=-\frac{\dot{H}}{H^2}\cong \frac{1}{2}\left(\frac{V_{\varphi}}{V} \right)^2$ is the slow roll parameter), it 
requires the inflaton's potential to be very flat and hence indicates a fine tuning issue \cite{Ijjas:2013vea}. In the simplest model of MBS, the amplitude of tensor fluctuation is 
comparable to that of curvature perturbation and thus the value of $r$ is too large \cite{Cai:2008qw}. Accordingly, one may expect certain dynamical mechanisms to be implemented in the 
MBS in order to enhance the amplitude of curvature perturbation to be consistent with the data, namely, a curvaton mechanism \cite{Cai:2011zx}. Also, it is possible to 
consider the effects of modified gravity theories such as from loop quantum cosmology to depress gravitational waves during the bounce \cite{deHaro:2014kxa}. Therefore, the no-detection 
of primordial gravitational waves can also impose a bound on various bounce models.

Fifth, a Universe whose background dynamics is realized by a primordial scalar field (not by isotropic fluids) has to reheat via decaying into light particles 
that will thermalize to match with the standard hot big bang expansion. Reheating could be produced due to the gravitational particle creation in 
an expanding Universe \cite{Parker:1968mv,Grib:1976pw}. In this case, an abrupt phase transition (a non-adiabatic transition) is needed in order to obtain sufficient 
particle creation that thermalizes, producing a reheating temperature that fits well with current observations. In bouncing cosmologies, the gravitational particle 
creation is natural to be implemented since the Universe would have experienced several phases including contracting, bouncing and expanding ones \cite{Quintin:2014oea}, and then, it 
is necessary to examine whether the reheating temperature is compatible with the current data \cite{Haro:2014wha}.

Sixth, bouncing cosmologies often suffer from a dangerous growth of primordial anisotropy of which the effective energy density scales as $a^{-6}$ in contracting phase. This is known 
as the famous Belinsky-Khalatnikov-Lifshitz (BKL) instability \cite{Belinsky:1970ew}. A solution to this problem can be realized in the ekpyrotic scenario \cite{Khoury:2001wf}, in 
which the energy density of the dominant matter scales with $a^{-q}$ with $q \gg 6$. To keep the 
scale invariance of primordial curvature perturbation, a model of MBS including a period of ekpyrotic contraction was recently suggested in \cite{Cai:2012va} and the 
evolution of primordial anisotropy was analyzed in \cite{Cai:2013vm}. However, it is important to be aware of this issue in other mechanisms of
nonsingular bounces, such as taking into account nonlinear matter contribution to smooth out the anisotropies \cite{Bozza:2009jx} or by including higher 
order curvature term \cite{Middleton:2008rh}.

Seventh, studies of distant type Ia supernovae \cite{Perlmutter:1998np} reveal that our Universe is expanding in an accelerating way. A viable cosmological model needs 
to accommodate with this late time acceleration, which usually is incorporated with a cosmological constant, or by a scalar field (see \cite{Copeland:2006wr} for recent reviews). There 
are other ways to implement the current cosmic acceleration, for example using $F({R})$ gravity (see for instance \cite{Nojiri:2006ri,Capozziello:2007ec}). Then it becomes important 
to question whether a bouncing solution can be realized in these different models \cite{Cai:2014jla}.

\vspace{0.25cm}

From the above considerations, the aim of the present work is to address, from a critical viewpoint, part of those points. In particular, dealing with a single scalar field, we 
provide a clear definition of the quasi-matter domination regime, which is the key to obtaining a running behavior of the spectral index, and we will see that this period is
characterized essentially by two {\it quasi-matter contraction} parameters. We perform a detailed calculation of the power spectrum of primordial perturbations, which allows us to 
obtain the expression of the spectral index and its running in terms of these two parameters associated with quasi-matter contraction. Once we have obtained these expressions, we can 
investigate the parameter spaces of some models, showing that these models of MBS could fit well with latest cosmological observations. We also present a detailed study of the 
gravitational reheating in the MBS when the background is depicted by holonomy corrected Loop Quantum Cosmology (LQC). More precisely,
 we show that when one considers a massless field nearly conformally coupled with gravity, the reheating temperature provided by this model is compatible with current observations. 
 Afterwards, we comment on several unclear issues that remain in specific models of MBS, such as nonlinearities produced during the bouncing phase, the Jeans instability in the case 
 of holonomy corrected LQC, as well the non-local Lorentz invariance of the scalar torsion in bouncing teleparallel models.

\vspace{0.25cm}

The units used throughout the paper are: $\hbar=c=8\pi G=1$.

\section{The phase of quasi-matter contraction}

In order to obtain a nearly matter dominated contracting phase, a simple way is to consider a potential of the form
\begin{eqnarray}\label{potential}
 V(\varphi) = V_0 e^{-\sqrt{3(1+\alpha)}|\varphi|} ~,
\end{eqnarray}
where $\alpha$ is a very small parameter. Then, the exact solution to the Friedmann and conservation equations,
\begin{eqnarray}
 \varphi(t) = -\frac{1}{\sqrt{3(1+\alpha)}} \ln \Big[ \frac{3(1+\alpha)^2}{2(1-\alpha)} V_0 t^2 \Big] ~,
\end{eqnarray}
depicts a quasi-matter dominated Universe with equation of state $P=\alpha\rho$. As we will see in next section, this background corresponds to a 
spectral index equal to $n_s-1=12\alpha$ with vanishing running.

To depict a viable phase of quasi-matter contraction, we consider, for a flat Friedmann-Lema\^{\i}tre-Robertson-Walker (FLRW) geometry, the Friedmann and 
conservation equations for the homogeneous part of a single scalar field,
\begin{eqnarray}\label{x1}
  H^2=\frac{1}{3}\left(\frac{\dot{\varphi}^2}{2}+V\right) ~; \quad \ddot{\varphi}+3H\dot{\varphi}+V_{\varphi}=0 ~.
\end{eqnarray}
Assuming that there was a quasi-matter domination in the contracting phase, i.e., $\dot{\varphi}^2\cong 2V \Longrightarrow \ddot{\varphi} \cong V_{\varphi}$, the above 
background equations become
\begin{eqnarray}\label{x2}
  \left\{\begin{array}{ccc}
          H^2&=& \frac{2}{3}V\\
          3H\dot{\varphi}+2V_{\varphi}&=&0
         \end{array}
\right.\Leftrightarrow
\left\{\begin{array}{ccc}
         {\mathcal H}^2&=& \frac{2}{3}a^2V\\
          3{\mathcal H}{\varphi}'+2a^2V_{\varphi}&=&0 ~,
         \end{array}
\right.
\end{eqnarray}
where ${\cal H}$ is the conformal Hubble parameter and a prime denotes a derivative with respect to conformal time.

This stage can be depicted by the equation of state parameter,
\begin{eqnarray}\label{x3}
 w\equiv\frac{P}{\rho}=
 %-1-\frac{2}{3}\frac{\dot{H}}{H^2}=
 -\frac{2}{3}\left( \frac{1}{2}+\frac{{\mathcal H}'}{{\mathcal H}^2} \right)
 \cong \frac{1}{3}\left(\frac{V_{\varphi}}{V} \right)^2-1 ~,
\end{eqnarray}
where $P$ and $\rho$ are the pressure and the energy density, which are related to the spectral index, and characterize this regime through the condition that $|w|\ll 1$.

Since a potential of the form $V_0e^{-\sqrt{{3}}|\varphi|}$ generates a phase of exact matter-domination, we would like to reexpress our potential $V$, for negative values 
of the field, as 
\begin{eqnarray}\label{potential1}
 V(\varphi)=V_0e^{\sqrt{{3}}\varphi(1+f(\varphi))} ~, 
\end{eqnarray}
and thus we obtain
\begin{eqnarray}\label{z1}
  w \cong {2}(\varphi f(\varphi))_{\varphi} ~,
\end{eqnarray}
where we have assumed that $|f(\varphi)|\ll 1$ and $|\varphi f_{\varphi}(\varphi)|\ll 1$. Further, it is convenient to introduce another parameter as follows,
\begin{eqnarray}\label{delta}
 \delta^2 \equiv \frac{w'}{{2}{\mathcal H}}\cong -\left(\frac{V_{\varphi}}{V} \right)_{\varphi} = -\sqrt{3}(\varphi f(\varphi))_{\varphi\varphi} \cong  -\frac{\sqrt{3}}{2}w_{\varphi} ~.
\end{eqnarray}
which is found to be related to the running of the spectral index in next section.

To demonstrate the small deviation from an exact matter-domination in the above model, we consider examples by choosing the 
functions $f(\varphi)=\frac{\beta\ln(-\varphi)}{\sqrt{3}\varphi}$ ($\beta>0$) and $f(\varphi)=\frac{\lambda^2}{\varphi^{2n}}$, for negatives values of the fields 
satisfying $|\varphi|\gg 1$. Then we obtain
\begin{eqnarray}
 w\cong \frac{2\beta}{\sqrt{3}\varphi} ~; \qquad w\cong -2(2n-1)\frac{\lambda^2}{\varphi^{2n}} ~,
\end{eqnarray}
and
\begin{eqnarray}
 \delta^2\cong \frac{\beta}{\varphi^2} ~; \qquad \delta^2\cong -\sqrt{3}2 n(2n-1)\frac{\lambda^2}{\varphi^{2n+1}} ~,
\end{eqnarray}
respectively.

Moreover, in analogue with inflationary cosmology, we can also define an effective number of e-folds before the end of the quasi-matter contracting 
phase as: $a(N)=e^{N}a_f$, where $a_f$ is the value of the scale factor at the end of this stage. With this definition, in the quasi-matter approximation, the number of e-folds is 
calculated as
\begin{eqnarray}
 N=-\int_{t_f}^{t_N}H(t)dt\cong \int_{\varphi(N)}^{\varphi_f}\frac{V}{V_{\varphi}}d\varphi ~,
\end{eqnarray}
which, in terms of the function $f(\varphi)$, can be expressed as,
\begin{eqnarray}
 N\cong \int_{\varphi(N)}^{\varphi_f}\frac{2}{\sqrt{3}(2+w)}d\varphi ~.
\end{eqnarray}

As a final remark to this section, we note that it is well-known that when one only considers a single canonical scalar field in spatially flat cosmologies, the frame of 
General Relativity (GR) forbids bounces from the contracting to the expanding phase, as is best seen by looking at the Raychaudhury equation in the flat 
FLRW space-time $\dot{H}=-\frac{1}{2}\dot{\varphi}^2<0$: the Hubble parameter always decreases, so it is absolutely impossible to 
pass from negative to positive values. For this reason, when the matter part of the Lagrangian is given in terms of a single scalar field, one is 
led to consider a non-canonical scalar field involving Horndeski operators such as model constructions in \cite{Cai:2012va, Cai:2013vm}, or to use 
cosmologies beyond the realm of GR, namely, holonomy corrected LQC, teleparallel $F(T)$ gravity, or $F(R)$ gravity.

%\vspace{0.5cm}

\section{Current MBS with a single scalar field: Background Review}

In this section we review the main MBS with a single scalar field in the frames of LQC and teleparallel $F(T)$ gravity, where the modified Friedmann equation is given by an 
ellipse in the plane $(H,\rho)$ that depicts a Universe moving clockwise along it \cite{Amoros:2013nxa}.

\subsection{Holonomy corrected LQC}

The main idea of LQC is that it assumes a discrete nature of space which leads to consider, at quantum level, a Hilbert space where quantum states 
are represented by almost periodic functions of the dynamical part of the connection \cite{Ashtekar:2003hd}. However, the connection variable does not correspond to a well 
defined quantum operator in such a Hilbert space and therefore the re-expression of the gravitational part of the Hamiltonian in terms of almost periodic functions (holonomies) is needed. 
It 
might be executed from a process of regularization. This new regularized Hamiltonian introduces a quadratic modification ($\rho^2$) in the Friedmann equation at high energies 
\cite{Singh:2006sg}, which can give rise to a nonsingular bounce when the energy density becomes equal to a critical value bellow the Planck energy density.
More precisely,
the holonomy corrected Friedmann equation in LQC, which depicts, in the plane $(H,\rho)$, the ellipse 
given by 
\begin{eqnarray}\label{friedmann}
 H^2 = \frac{\rho}{3}\left(1-\frac{\rho}{\rho_c}\right) ~,
\end{eqnarray}
where $\rho_c$ is the so-called ``critical density'' (the energy at which the Universe bounces).

\subsection{Teleparallelism}

It was also noticed in \cite{Cai:2011tc} that a nonsingular bouncing solution can be derived in theories of teleparallel gravity. Teleparallel theories are
based on the {\it Weitzenb\"ock} space-time. This space is ${\R}^4$, with a Lorentz metric, in which a global, orthonormal basis of its tangent bundle given by 
four vector fields $\{ e_i \}$ has been selected, that is, they satisfy $g(e_i, e_j)= \eta_{ij}$ with $\eta=\mbox{diag}\,(-1,1,1,1)$. The Weitzenb\"ock 
connection $\nabla$ is defined by imposing that the basis vectors $e_i$ be absolutely parallel, i.e., $\nabla e_i=0$.

The Weitzenb\"ock connection is compatible with the metric $g$, and it has zero curvature because of the global parallel transport defined 
by the basis $\{ e_i \}$. The information of the Weitzenb\"ock connection is carried by its torsion, and its basic invariant is the scalar 
torsion ${T}$. The connection, and its torsion, depend on the choice of orthonormal basis $\{ e_i \}$, but if one adopts the flat Friedmann-Lema\^{\i}tre-Robertson-Walker 
(FLRW) metric and selects as orthonormal basis $\{ e_0= \partial_0, e_1= \frac{1}{a} \partial_1, e_2= \frac{1}{a} \partial_2, e_3= \frac{1}{a} \partial_3 \}$, then the scalar 
torsion is ${T}=-6H^2$, where $H=\frac{\dot{a}}{a}$ is the Hubble parameter, and this identity is invariant with respect to local Lorentz transformations that only depend 
on time, i.e. of the form $\tilde{e_i}= \Lambda^k_i(t) e_k$ (see \cite{Bengochea:2008gz,deHaro:2012zt}).

\bigskip

With the above choice of orthonormal fields, the Lagrangian of the $F({T})$ theory of gravity is expressed as:
\begin{eqnarray}\label{eq:lagr}
 {\mathcal L}_{T}={\mathcal V}(F({ T})+{\mathcal L}_M) ~,
\end{eqnarray}
where, once again, ${\mathcal V}=a^3$ is the element of volume, and ${\mathcal L}_M$ is the matter Lagrangian density. Accordingly, the Hamiltonian of the system is given by
\begin{eqnarray}\label{eq:ham}
 {\mathcal H}_{T}= \left(2T\frac{dF({T})}{dT}-F({ T})  +\rho \right){\mathcal V} ~,
\end{eqnarray}
where ${\rho}$ is the energy density. Imposing the Hamiltonian constrain ${\mathcal H}_{T}=0$ leads to the modified Friedmann equation
\begin{eqnarray}\label{eq:friedmann}
\rho=-2 \frac{dF({ T})}{dT}{T}+F({T})\equiv G({T}) ~,
\end{eqnarray}
which defines a curve in the plane $(H,\rho)$ by applying the relation $T=-6H^2$. It is interesting to note that Eq. (\ref{eq:friedmann}) may be inverted, and 
therefore, each curve of the form $\rho=G({T})$ defines an $F({T})$ theory through the following expression:
\begin{eqnarray}\label{eq:F(T)}
 F({ T})=-\frac{\sqrt{-{ T}}}{2}\int \frac{G({ T})}{{T}\sqrt{-{ T}}}d{ T}.
\end{eqnarray}

\bigskip

In order to produce a cyclically evolving Universe, let us take the $F({T})$ theory arising from the ellipse that defines the holonomy corrected Friedmann 
equation in LQC (Eq. (\ref{friedmann})). To obtain a parametrization of the form $\rho=G({T})$, the curve has to be split in two branches
\begin{eqnarray}\label{eq:Gpm}
 \rho=G_{\pm}({ T})=\frac{\rho_c}{2}\left(1\pm\sqrt{1+\frac{2{ T}}{\rho_c}}\right),
\end{eqnarray}
where the branch $\rho=G_-({ T})$ corresponds to $\dot{H}<0$ and $\rho=G_+({ T})$ is the branch with $\dot{H}>0$. Applying Eq. (\ref{eq:F(T)}) to these branches 
yields (\cite{Amoros:2013nxa})
\begin{eqnarray}\label{eq:Fpm}
 F_{\pm}({ T})=\pm\sqrt{-\frac{{ T}\rho_c}{2}}\arcsin\left(\sqrt{-\frac{2{T}}{\rho_c}}\right)+G_{\pm}({ T}) ~.
\end{eqnarray}

\subsection{Phase space dynamics}

In holonomy corrected LQC and teleparallel $F(T)$ gravity with a single scalar field, namely $\varphi$, the dynamical equation of the background is given by the conservation equation
\begin{eqnarray}\label{KG}
 \ddot{{\varphi}}+3{H}_\pm\dot{{\varphi}}+{V}_{\varphi}(\varphi)=0 ~,
\end{eqnarray}
where ${H}_- =-\sqrt{\frac{{\rho}}{3}(1-\frac{\rho}{\rho_c})}$ in the contracting phase, ${H}_+ = \sqrt{\frac{{\rho}}{3} (1-\frac{\rho}{\rho_c})}$ in the expanding one.

The goal of this Section is to prove that, when one deals with a potential that defines a quasi-matter domination at early times in the contracting phase, i.e., with the form
$V(\varphi)=V_0 e^{\sqrt{3}\varphi(1+f(\varphi))}$, all the solutions of Eq. (\ref{KG})  depict, at early times, a matter dominated Universe.

%When one deals with a potential that defines a quasi-matter domination at early times in the contracting phase, i.e., with the form
%$V(\varphi)=V_0 e^{\sqrt{3}\varphi(1+f(\varphi))}$, 

To show this property, first of all, note that
Eq. (\ref{KG}) has a solution with the following asymptotic behavior at early times
\begin{eqnarray}\label{sol}
 \varphi(t)\rightarrow -\frac{2}{\sqrt{3}}\ln\left(-\sqrt{\frac{3}{2}V_0}t\right) ~, \quad \mbox{ when } \quad t\rightarrow -\infty,
\end{eqnarray}
because one can check that $-\frac{2}{\sqrt{3}}\ln\left(-\sqrt{\frac{3}{2}V_0}t\right)$ is solution of the equation
\begin{eqnarray}
 \ddot{{\varphi}}+3{H}_\pm\dot{{\varphi}}+{V}_{\varphi}(\varphi)=0 ~,
\end{eqnarray}
with $H_ -=-\sqrt{\frac{\rho}{3}}$ and $V(\varphi)=V_0 e^{\sqrt{3}\varphi}$.

%Then, our statement  will be proved if we show  that the all the solutions of (\ref{KG})  have the asymptotic behavior depicted in (\ref{sol}).

%The method to prove this asymptotical behavior goes as follows:
Secondly, 
at early and late times one can disregard holonomy corrections, and we can reduce the analysis to large values of $|\varphi|$, and thus, take the 
potential ${V}(\varphi)=V_0e^{\sqrt{3}\varphi}$. Then, performing the change of variable ${\varphi} = -\frac{2}{\sqrt{3}} \ln \psi$, the corresponding non-perturbed 
Klein-Gordon equation (or equivalently, the conservation equation) reads
\begin{eqnarray}\label{eq1}
 \frac{d\dot{\psi}}{d{\varphi}}=F_{\pm}(\dot{\psi}) ~,
\end{eqnarray}
with
\begin{eqnarray}
 F_{\pm}(\dot{\psi}) = \frac{3\sqrt{3}}{4\dot{\psi}} \left(\frac{2}{3}\dot{\psi}^2 +V_0 \right) \mp\frac{3}{2} \sqrt{\frac{2}{3}\dot{\psi}^2+V_0} ~,
\end{eqnarray}
where in $F_{\pm}$, the sign $+$ (resp. $-$) means that the Universe is in the expanding (resp. contracting) phase. Note that,
Eq. (\ref{eq1}) describes two (one for the contracting and other one for the expanding phase) one-dimensional first order autonomous dynamical systems, that 
are completely understood calculating critical points. These critical points are $\dot{\psi}_{+}=\sqrt{\frac{3}{2}V_0}$ for the expanding phase and 
$\dot{\psi}_{-}=-\sqrt{\frac{3}{2}V_0}$ for the contracting one,
%Taking into account the sign of $F_{\pm}$, one deduces the dynamics of the system in
%time $\varphi$ (in figure $1$ we show the phase portrait (in the straight line $\dot{\psi}$) when the Universe is in the contracting phase, which is the case 
%we are interested in), and from the phase portrait in time $\varphi$, we deduce the corresponding one in cosmic time, changing the direction of the arrows 
%for negative values of $\dot{\psi}$ (see figure 2).
meaning that in the  contracting phase  $\dot{\psi}_{-}$ is a global repeller, i.e., at very early time all the solutions have the asymptotic behavior 
$\dot{\psi}(t)\rightarrow \dot{\psi}_{-}$ when $t\rightarrow -\infty$. 

Finally, the point $\dot{\psi}_{-}$  corresponds to the solution 
$\varphi = -\frac{2}{\sqrt{3}} \ln\left(-\sqrt{\frac{3}{2}V_0}t\right)$, which satisfies $\rho = \dot{\varphi}^2 =\frac{4}{3t^2}$, what proves that all the solution
of Eq. (\ref{KG}) depict, at early times, a matter-dominated the Universe  in the contracting phase.

\section{Detailed calculation of the power spectrum in quasi-matter bounce}

Common to all these cases is the Mukhanov-Sasaki \cite{Mukhanov:1985rz} equation for scalar perturbations, in Fourier space, which can be expressed as
\begin{eqnarray}\label{MS}
 v_k''+\left(c_s^2k^2-\frac{z''}{z}\right)v_k=0 ~,
\end{eqnarray}
where for very low energy densities and curvatures, $c_s^2=1$ and $z=a\frac{\dot{\varphi}}{H}=a\frac{{\varphi}'}{{\mathcal H}}$. The explicit expressions for 
$z$ and $c_s^2$ in the cases of $F(T)$ and $F(R)$ gravities were derived in \cite{Haro:2013bea} and \cite{Hwang:1996bc}, respectively. In contrast, 
for the holonomy corrected LQC, the form of $z$ is the same as in GR, i.e., $z=a\frac{{\varphi}'}{{\mathcal H}}$. However, the expression for the square of 
the sound speed parameter differs from unity at high energies \cite{Cailleteau:2011kr}. As we will discuss later, in the holonomy corrected LQC, the value 
of the square of the sound speed parameter becomes negative during the super-inflationary phase $\dot{H}>0$, which happens when the energy density is greater than
one half the critical density.

During the quasi-matter dominated contraction, which occurs in the contracting phase at low energies and curvatures, one has 
$\frac{z''}{z}\cong \frac{a''}{a}$ (a detailed derivation of this relation was done in \cite{Elizalde:2014uba}), which means
\begin{eqnarray}\label{x7}
 \frac{z''}{z} \cong {\mathcal H}' +{\mathcal H}^2 ~.
\end{eqnarray}
Accordingly, solving Eq. (\ref{x3}) for a constant $w$, i.e., taking ${\mathcal H} \cong \frac{2}{\eta}(1-{}{3}w)$ and inserting into (\ref{x7}), we get
\begin{eqnarray}
 \frac{z''}{z} \cong \frac{2}{\eta^2}(1-{}{9}w) ~,
\end{eqnarray}
under the assumption of $|w|\ll1$. We refer to \cite{Cai:2009hc} for a more generic analysis.

It is clear from this result that the Mukhanov-Sasaki equation (\ref{MS}) during the quasi-matter dominated contraction can be approximately expressed as
\begin{eqnarray}\label{x10}
  v_k''+\left[ k^2 - \frac{1}{\eta^2}\left(\nu^2-\frac{1}{4} \right) \right]v_k=0; \quad \nu\cong \frac{3}{2}-{}{6}w ~.
\end{eqnarray}
Assuming the initial conditions of primordial perturbations to be vacuum fluctuations, one then obtains the solution to Eq. (\ref{x10})
\begin{eqnarray}\label{hankel}
 v_k=\frac{\sqrt{\pi|\eta|}}{2}e^{i(1+2\nu)\frac{\pi}{4}}H^{(1)}_{\nu}(k|\eta|) ~.
\end{eqnarray}
For modes well outside of the Hubble radius $k|\eta|\ll 1$,  (\ref{x10}) becomes
\begin{eqnarray}\label{x12}
 v_k''- \frac{1}{\eta^2}\left(\nu^2-\frac{1}{4} \right)v_k=0 ~,
\end{eqnarray}
and then the dominant mode of the solution is given by
\begin{eqnarray}\label{y13}
 v_k \cong C_2(k)|\eta|^{\frac{1}{2}-\nu} ~.
\end{eqnarray}

On the other hand, to calculate the power spectrum we will choose a pivot scale, namely $k_*$, and let $\eta_*$ be the time  at which the pivot scale crosses the Hubble radius 
in the contracting phase. Then we could write the scale factor as follows
\begin{eqnarray}\label{A1}
 a(\eta) = a_*\left(\frac{\eta}{\eta_*}\right)^{\frac{1}{2}+\nu} \cong \frac{k_*}{|H_*|} \left(\frac{k_*|\eta|}{2}\right)^{\frac{1}{2}+\nu} ~,
\end{eqnarray}
where $a_*$ and $H_*$ are the values of the scale factor and Hubble parameter, respectively, at the crossing time. The approximation on the r.h.s.  comes from the fact that 
in the quasi-matter dominated contracting phase, we have $aH\cong \frac{2}{\eta}$. Hence, since in the quasi-matter domination one has 
$z(\eta)=\sqrt{3(1+w)}a(\eta)\cong \sqrt{3}a(\eta)$, the solution (\ref{y13}) can be written as follows,
\begin{eqnarray}\label{zz1}
 v_k \cong {3\sqrt{3}}\frac{k_*}{|H_*|} |\eta_*|^{-\frac{1}{2}-\nu} {C}_2(k) \left(z(\eta)\int_{-\infty}^{\eta} \frac{d\bar\eta}{z^2(\bar\eta)}\right) ~.
\end{eqnarray}

For modes well outside of the Hubble radius, the solution (\ref{hankel}) should match Eq. (\ref{zz1}). Using the small argument approximation in the Hankel function and 
expression (\ref{y13}), we find
\begin{eqnarray}\label{Zeta}
 \xi_k(\eta)\equiv\frac{v_k}{z(\eta)}\cong -\frac{i}{16}\left(\frac{6}{k}\right)^{\frac{3}{2}}e^{i(1+2\nu)\frac{\pi}{4}}
 \frac{k_*^3}{|H_*|} \int_{-\infty}^{\eta} \frac{d\bar\eta}{z^2(\bar\eta)} \left(\frac{k}{k_*} \right)^{\frac{3}{2}-\nu}.
\end{eqnarray}
These modes will re-enter the Hubble radius at late times in the expanding phase, when the Universe will be matter dominated. Then, the power spectrum is given by
\begin{eqnarray}\label{aa1}
 {\mathcal P}_{\xi}(k)=\frac{27}{64\pi^2} \frac{k_*^6}{H_*^2} \left(\int_{-\infty}^{\eta} \frac{d\bar\eta}{z^2(\bar\eta)}\right)^2 \left(\frac{k}{k_*} \right)^{3-2\nu} ~.
\end{eqnarray}

Provided that the Universe is always dominated by the same matter field throughout the contracting and expanding phases, 
and assuming that the  usual procedures apply all through the non-singular bouncing phase, which is justified due to the analycity of $\xi_k(\eta)$ (see Eq. (\ref{Zeta}))
in the bouncing phase,
one can evaluate the above equation and eventually obtain 
the final formula for the power spectrum corresponding to scalar perturbations at the pivot scale $k_*$:
\begin{eqnarray}\label{power}
 {\mathcal P}_{\xi}(k_*) = \frac{27}{64\pi^2}\frac{k_*^6}{H_*^2} \left(\int_{-\infty}^{+\infty} \frac{d\eta}{z^2(\eta)}\right)^2 ~,
\end{eqnarray}
at late times of cosmic evolution. 

The calculation of the power spectrum, in general, can be performed numerically as pointed out in \cite{deHaro:2014kxa}. However, dealing with LQC or teleparallel $F(T)$ gravity, in 
the simple case of a matter dominated Universe throughout the whole background evolution, that is, for the scale factor 
$a(t) = a_*\left(\frac{9}{4}H_*^2t^2+\frac{3H_*^2}{\rho_c} \right)^{1/3}$, where $\rho_c$ is once again the critical density, i.e., the value of the energy density at the 
bouncing time, the calculation could be done analytically giving as a result
\begin{eqnarray}\label{A}
 {\mathcal P}_{\xi}(k_*)=\frac{\pi^2}{9}\frac{\rho_c}{\rho_{pl}} ~,
\end{eqnarray}
in the case of holonomy corrected LQC \cite{WilsonEwing:2012pu}, and
\begin{eqnarray}\label{B}
 {\mathcal P}_{\xi}(k_*) =\frac{16}{9}\frac{\rho_c}{\rho_{pl}}{\mathcal C }^2 ~,
\end{eqnarray}
where ${\mathcal C} \cong 0.9159$ is Catalan's constant and $\rho_{pl}$ is the Planck density, for the case of teleparallel $F(T)$ gravity \cite{Haro:2013bea}.

It is important to note that in the above calculation we have assumed that the Universe is dominated by a single matter field throughout the whole primordial epoch.
However, a much more realistic Universe has to involve other matter components, such as radiation and a cosmological constant \cite{Cai:2014jla}. In addition, one
often introduce an ekpyrotic matter field to dilute unnecessary anisotropies generated in contracting phase \cite{Cai:2014zga}. In these cases, the amplitude of 
primordial curvature perturbation will be different and it is possible to obtain the correct amplitude which satisfies the CMB observations even when $\rho_c$ is of 
the order of the Planck energy density \cite{Cai:2014jla}.

%\begin{remark}
 In order for the theoretical results provided by both models (Eqs. (\ref{A}) and (\ref{B})) to match with the CMB data ${\mathcal P}_{\xi}\cong 2\times 10^{-9}$, the value of 
 the energy density at the bouncing point has to be at least of the order $\rho_c\sim 10^{-9}\rho_{pl}$ or even larger.
%\end{remark}

From Eq. (\ref{aa1}) and following the definition of the spectral index for scalar perturbations, one can derive the following expression:
\begin{eqnarray}
 n_s-1\equiv\frac{d\ln {\mathcal P}_{\xi}(k)}{d\ln k}=3-2\nu={}{12}w ~.
\end{eqnarray}
Note that one can further expand the spectral index by involving a running index via the parametrization, $n_s\cong n_s(k_*)+\alpha_s(k_*)\ln\left(\frac{k}{k_*}\right)$. The 
running of the spectral index, $\alpha_s$, is then obtained as
\begin{eqnarray}
 \alpha_s\equiv \left(\frac{dn_s}{d\ln k}\right)_{k=aH} = -\frac{2n_s'}{(1+3w){\mathcal H}}\cong -\frac{{}{24}w'}{{\mathcal H}}(1-3w)\cong-48\delta^2 ~,
\end{eqnarray}
where we have used Eqs. (\ref{x3}) and (\ref{delta}).

\begin{remark}
The quasi-matter domination was first addressed in \cite{Cai:2014jla} in the context of the $\Lambda$CDM model. In this context
\begin{eqnarray}
 w=\frac{-\rho_{\Lambda}}{\rho_{\Lambda}+\rho_{CDM}}<0,
\end{eqnarray}
where $\rho_{\Lambda}$ is the energy density corresponding to the cosmological constant $\Lambda$ and $\rho_{CDM}$ is the energy density of the Cold Dark Matter (CDM). When the CDM 
dominates the evolution of the Universe, the value of $w$ is negative and close to zero and it is easy to show that the spectral index is given by $n_s-1=12w$. In this 
specific case, after applying the approximate relations $w' = 3{\mathcal H}(w+w^2) \cong 3{\mathcal H}w$ and $\frac{{\mathcal H}'}{{\mathcal H}^2}=-\frac{1}{2}\left({3}w+{2} \right)$, one 
can calculate the running of the spectral index and finds
\begin{eqnarray}
 \alpha_s \equiv \left(\frac{dn_s}{d\ln k}\right)_{k=aH} = \frac{12w'{\mathcal H}}{ {\mathcal H}'} \cong -72 w >0 ~.
\end{eqnarray}
As a result, the $\Lambda$CDM bounce model predicts a positive running behavior \cite{Cai:2014jla}, which is not ruled out yet 
by observations. Note also that the same happens if one mixes CDM with an isotropic fluid with negative pressure $P = w \rho$ ($w<0)$.
\end{remark}

In the same way, for tensor perturbations one has $\frac{a''}{a} = \mathcal{H}' +\mathcal{H}^2 \cong \frac{2}{\eta^2}(1-9w)$, and thus, the following power spectrum is obtained:
\begin{eqnarray}
 {\mathcal P}_T(k) = \frac{32}{81\pi^2}\frac{k_*^6}{H_*^2} \left(\int_{-\infty}^{+\infty} \frac{d\eta}{z_T^2(\eta)} \right)^2 \left(\frac{k}{k_*} \right)^{3-2\nu} ~,
\end{eqnarray}
where, for very low energy densities and curvatures, $z_T=a$. The exact expression of $z_T$ in holonomy corrected LQC was obtained in \cite{Cailleteau:2012fy}, in 
teleparallel $F(T)$ gravity in \cite{Haro:2013bea}, and in modified $F(R)$ gravity in \cite{Hwang:1996xh}.

In the MBS, the spectral index for tensor perturbation and its running are
\begin{eqnarray}
 n_T\equiv \frac{d{\mathcal P}_T(k) }{d\ln k} = n_s-1, \quad \alpha_T\equiv \left(\frac{dn_T}{d\ln k}\right)_{k=aH} =\alpha_s ~,
\end{eqnarray}
and the ratio of tensor-to-scalar perturbations, is given by
\begin{eqnarray}\label{ratio}
 r \equiv \frac{{\mathcal P}_T(k)}{{\mathcal P}_{\xi}(k)} = 
 \frac{8}{3} \left(\frac{\int_{-\infty}^{+\infty} {z_T^{-2}(\eta)}d\eta}{\int_{-\infty}^{+\infty} {z^{-2}(\eta)}d\eta} \right)^2 ~.
\end{eqnarray}

\section{Fitting the spectral index and the running parameters}

The Planck2013 data released in 2013 constrain the values of the spectral index and its running to be 
$$ n_s=0.9603\pm 0.0073~,~~~ \alpha_s=-0.0134\pm 0.0090~,$$
at $1\sigma$ \cite{Ade:2013zuv}. In models of slow roll inflation (such as monomial, natural, hilltop and plateau potentials) and in the MBS with a quasi-matter domination, in 
general, $1-n_s$ is of order $N^{-1}$ ($N$ being the number of e-folds before the end of the corresponding regime), while the running parameter is of order $N^{-2}$ and, 
consequently, one has $\alpha_s\sim (1-n_s)^2$. This situation is compatible with the Planck2013 data. For example, in the case of the MBS, for large negative values of 
$\varphi$, with the potential $V(\varphi) = e^{\sqrt{3}\varphi} (-\varphi)^{\beta}$ (i.e., for $f(\varphi)=\frac{\beta\ln(-\varphi)}{\sqrt{3}\varphi}$), one has
\begin{eqnarray}\label{zzz}
 n_s-1=\frac{24\beta}{\sqrt{3}\varphi},\quad \alpha_s=-\frac{48\beta}{\varphi^2},
\end{eqnarray}
which means that $\alpha_s=-\frac{1}{4\beta}(1-n_s)^2$. This situation is compatible with a large range for $\beta>0$. Choosing for instance $\beta=\frac{1}{24}$, the 
constrains (\ref{zzz}) are compatible with the Planck2013 data for $$\varphi\in(-17.81,-12.28).$$

Dealing with the MBS model $f(\varphi)=\frac{1}{\varphi^{2}}$, as has been introduced in the first section, one has
\begin{eqnarray}\label{xxx}
 n_s-1=-\frac{24}{\varphi^2},\quad \alpha_s=\frac{96\sqrt{3}}{\varphi^3} ~,
\end{eqnarray}
which leads to the relation $\alpha_s^2=2(1-n_s)^3$. A simple calculation shows that (\ref{xxx}) fits well the Planck2013 data for values of the field in the range
\begin{eqnarray}
 \varphi\in (-27.24,-22.60) ~.
\end{eqnarray}

The same happens, in MBS,  with  exponential potentials of the form $V=V_0\frac{e^{-\sqrt{3}|\varphi|}}{1+e^{-\beta |\varphi|}}$ with $\beta>0$. In this case one has
\begin{eqnarray}
 n_s-1=-\frac{24\beta}{\sqrt{3}}e^{-\beta |\varphi|}, \quad \alpha_s=-48\beta^2e^{-\beta |\varphi|} ~,
\end{eqnarray}
which lead to the relation $\alpha_s=2\sqrt{3}\beta (n_s-1)$. This relation is compatible with the Planck2013 data for a wide range of values of $\beta$, for example, 
when $\beta=\frac{1}{8\sqrt{3}}$.

\vspace{0.25cm}

On the other hand, in slow roll inflation one also has to take into account the latest CMB constrain ${r}<0.12$, and the fact that more than $50$ e-folds are needed to solve 
the horizon and flatness problems of standard big bang cosmology. By studying the $(n_s,r)$ plane, the Planck2013 data is able to constrain the parameter space of 
inflation models very well, preferring potentials with a concave shape ($V_{\varphi\varphi}<0$) \cite{Ade:2013uln}. Moreover, these models could be further 
constrained if the data of the running parameter $\alpha_s$ were included \cite{Easther:2006tv}.

Effectively, in slow roll inflation, (see for instance \cite{Bassett:2005xm} for a review of inflationary cosmology) the commonly used parameters are:
\begin{eqnarray}\label{x14}
 \bar\epsilon= -\frac{\dot{H}}{H^2}\cong \frac{1}{2}\left(\frac{V_{\varphi}}{V} \right)^2, \quad \bar\eta=2\epsilon-\frac{\dot{\epsilon}}{2H\epsilon}
\cong\frac{V_{\varphi\varphi}}{V} ~,
\end{eqnarray}
and the spectral index  and its running are given by
\begin{eqnarray}\label{x15}
 {n}_s-1=2\bar\eta-6\bar\epsilon,\quad {\alpha}_s=16\bar\epsilon\bar\eta-24\bar\epsilon^2-2\bar\xi,
\end{eqnarray}
where we have introduced the second order slow roll parameter
\begin{eqnarray}
 \bar\xi\equiv \left(2\bar\epsilon-\frac{\bar\eta'}{{\mathcal H}\bar\eta}\right)
 \bar\eta \cong \frac{V_{\varphi}V_{\varphi\varphi\varphi}}{V^2}.
\end{eqnarray}
Moreover, in inflationary cosmology, the tensor-to-scalar ratio $r$, is related to the slow roll parameter $\bar\epsilon$, in the way
\begin{eqnarray}\label{x18}
 {r}=16\bar\epsilon,
\end{eqnarray}
which does not happen in the MBS.

Dealing with the running, a simple calculation leads to the relation
\begin{eqnarray}
 \alpha_s=\frac{1}{2}(n_s-1)r+\frac{3}{32}r^2-2\bar{\xi} ~.
\end{eqnarray}
Then, choosing for instance,  the $\Lambda$CDM$+r+\alpha_s$ model from Planck2013 combined with WP and BAO data, which gives the following results
$n_s=0.9607\pm 0.0063$, $r\leq 0.25$ at $95$\% C.L. and $\alpha_s=-0.021^{+0.012}_{-0.010}$ (see table 5  of $\cite{Ade:2013uln}$).
We will consider $n_s$ at 2$\sigma$ C.L. and take the conservative bound $r\leq 0.32$ (see figure 4 of $\cite{Ade:2013uln}$). Since   the minimum of the function
$ \frac{1}{2}(n_s-1)r+\frac{3}{32}r^2$ is reached at the point ($n_s=0.9481, r=0.1384$)
given a value  greater than $-0.0018$, one obtains  the bound
\begin{eqnarray}\label{X}
\alpha_s\geq -0.0018-2\bar\xi,
\end{eqnarray}
meaning that plateau potentials such as $V(\varphi)=V_0\left(1-\frac{\varphi^2}{\mu^2}\right)$
(Hill-Top Inflation (HTP)) \cite{Boubekeur:2005zm}, $V(\varphi)=V_0\left(1-\frac{\varphi^2}{\mu^2}\right)^2$ (Double-Well Inflation (DWI)) \cite{Olive:1989nu},
$V(\varphi)=V_0\left(1+\cos\left(\frac{\varphi}{\mu}\right)\right)$ (Natural Inflation (NI)) \cite{Freese:1990rb}, or
$V(\varphi)=V_0\left(1+\alpha\ln\left(\cos\left(\frac{\varphi}{\mu}\right)\right)\right)$ (Pseudo Natural Inflation (PSNI)) \cite{randall},
when one considers values of the running at 1$\sigma$ C.L.,
are disfavoured
by Planck data because for all of them $\bar\xi\leq 0$. In fact, the distance from the theoretical value of the runnig to its expected observable value, namely ${\frak D}$, is larger 
than
$1.6\sigma$.

%and, in modulus, the theoretical values given by these models are always lower than  the expected observational one,
%what means that, from the Planck2013 data, these models are disfavored at 94.5\% C.L.

A distance larger than $1.6\sigma$ is also obtained for
the potential that leads to Exponential SUSY Inflation (ESI) $V(\varphi)=V_0\left(1-e^{-p\varphi}\right)$ \cite{stewart}, for Power Law Inflation (PLI) 
whose potential is given by $V(\varphi)=V_0e^{-p\varphi}$ \cite{lucchin}, 
for  K\"aller Moduli Inflation I (KMII) \cite{conlon} given by the potential $V(\varphi)=V_0\left(1-\alpha\varphi e^{-\varphi}\right)$, 
in Witten-O'Raifeartaigh Inflation (WRI) \cite{Witten} with potential
$V(\varphi)=V_0\ln^2\left(\frac{\varphi}{\mu} \right)$,
and for general hill-top potentials such as: $V(\varphi)=V_0\left(1-\left(\frac{\varphi}{\mu}\right)^p\right)$ (Small Field Inflation (SFI))  \cite{albrecht} and
$V(\varphi)=V_0\left(1+\left(\frac{\varphi}{\mu}\right)^p\right)$
(Valley Hybrid Inflation (VHI))  \cite{LINDE}
with $p\geq 3$,
because in these
cases one has $\bar\xi\leq \bar\eta^2$, that is,
\begin{eqnarray}
\alpha_s\geq \frac{r}{8}\left((n_s-1)+\frac{3}{16}r \right)-\frac{1}{2}(n_s-1)^2=-\frac{3r^2}{32}\geq -0.0018,
\end{eqnarray}
where we have evaluated $\alpha_s$, as a function of $n_s$ and $r$  at the absolute minimum, namely $n_s=1-\frac{3r}{8}$ with $r=0.1384$, in the 
rectangle $ [0.9481, 0.9733]\times [0,0.32]$.

For potentials such as:
$V(\varphi)=V_0\left(1-\left(\frac{\varphi}{\mu}\right)^{-p}\right)$ (Brane Inflation (BI)) \cite{espinosa},
and $V(\varphi)=V_0\left(1+\left(\frac{\varphi}{\mu}\right)^{-p}\right)$ (Dynamical Supersymmetric Inflation (DSI))  \cite{casas},  since one has
$\bar\xi=\frac{p+2}{p+1}\bar\eta^2$, 
%one can obtain
%the following exact formula
%\begin{eqnarray}\label{AA}
%\alpha_s=\frac{(p-2)r}{8(p+1)}\left((n_s-1)+\frac{3r}{16}\right)
%\nonumber\\
%-\frac{p+2}{2(p+1)}(n_s-1)^2.
%\end{eqnarray}
and
the minimum of $\alpha_s$ is obtained at $n_s=1-\frac{3r}{8}$ with $r=0.1384$, once again, one gets
%Using that $\frac{p-2}{p+1}\leq 1$ and $\frac{p+2}{p+1}\leq \frac{3}{2}$, one has at the point ($n_s=0.9481,r=0.138$) the following bound
\begin{eqnarray}
\alpha_s\geq-\frac{3r^2}{32}\geq -0.0018,
\end{eqnarray}
which is also incompatible with the runnig provided by Planck2013 at $1\sigma$ C.L., because
${\frak D}\geq 1.6\sigma$.

Finally, dealing with  Large Field Inflation (LFI) given by the monomial  potential $V(\varphi)=V_0\varphi^{p}$ \cite{Lindea}, and 
Radiation Gauge Inflation (RGI) \cite{honorez},  whose potential is given by $V(\varphi)=V_0\frac{\varphi^2}{\alpha+\varphi^2}$, one has
\begin{eqnarray}
 \alpha_s\geq -\frac{2}{3}(n_s-1)^2\geq -0.0018,
\end{eqnarray}
where we  have evaluated this quantity at $n_s=0.9481$, giving ${\frak D}\geq 1.6\sigma$.

Numerically we have tested all the  $49$ potentials  that appear in table 1 of \cite{MARTIN}: Using 
Planck2013+WP+BAO:$\Lambda$CDM+$r$+$\alpha_s$ data,
 $45$ of the potentials in the list of \cite{MARTIN} obtaining ${\frak D}\geq 1.6\sigma$. The other four potentials, namely Loop Inflation \cite{binetruy}, Radiatively 
 Corrected Higgs 
 Inflation \cite{Barvinsky},
$\beta$ exponential inflation \cite{Alcaniz} and  Generalized Mixed Inflation \cite{Kinney},  ${\frak D}\geq 1.4\sigma$.
More restictive are the data provided 
Planck2013+WP+high-${{\ell}}$:$\Lambda$CDM+$r$+$\alpha_s$ (see table 5 of \cite{Ade:2013uln}) where all of the $49$ models ${\frak D}$ is greater than $1.6\sigma$.

From the results provided by Planck2013 and these analyses, one could find that, the realization of a sizable negative running behavior implies that inflationary cosmology may 
require multiple 
fields or a breakdown of slow roll 
condition. For instance, following the second path, in \cite{Cai:2014vua} a short violation of the slow roll condition has been considered, due to an appearance of a step-like potential 
from the perspective of string theory.
Fortunately, for single field slow roll inflation, the new Planck2015 observational data \cite{Planck:2015xua,Planck:2015xub}, reduce the modulus of the running one order what allows the viablility
of the majority of single scalar field slow roll inflationary models. For example, using
Planck2015+TT+lowP:$\Lambda$CDM+$r$+$\alpha_s$, where $n_s=0.9667\pm 0.0066$ 
and $\alpha_s=-0.0126^{+0.0098}_{-0.0087}$
at $1\sigma$ C.L. (see table $4$ of \cite{Planck:2015xub}), and the conservative constrain $r\leq 0.25$ 
(see figure $6$ of \cite{Planck:2015xub}), 
%and $\alpha_s=-0.0126^{+0.0098}_{-0.0087}$ (formula (42a) of \cite{Planck:2015xua}), 
for all the potentials depicted above one has ${\frak D}\cong 1.1\sigma$. Moreover, if one introduces lensing, then ${\frak D}$ will be reduced to be lower than $1\sigma$, and thus allowing 
single scalar field slow roll 
inflation.

A final remark to end this section is in order:
The same kind of results are obtained in inflation (resp. MBS), when one deals with  an equation of state $P= -\rho + \frac{\beta}{(N+1)^{\alpha}} \rho$ \cite{Mukhanov:2013tua} 
for inflation (resp. $P=\frac{\beta}{(N+1)^{\alpha}} \rho$ in MBS) that depends on the number of e-folds $N$ before the end of the inflationary (resp. quasi-matter domination) period. 
In fact, for theMBS one has
\begin{eqnarray}\label{alpha_s}
 \alpha_s=\frac{2\alpha}{N+1}(n_s-1) ~,
\end{eqnarray}
which is perfectly compatible with current observation data, taking for example $\alpha=12$, $\beta=-\frac{1}{2}$, and $N=12$ (see for instance \cite{Elizalde:2014uba}). Recall 
that in bouncing cosmologies a large number of e-folds is {\it not} required, since at the bounce point all parts of the Universe are already in causal contact, and also the 
flatness problem is avoided, because the contribution of the spatial curvature decreases in the contracting phase at the same rate as it increases in the expanding 
one (see for instance \cite{Brandenberger:2012uj}). However, in inflation, in order to solve the horizon and flatness problems, the number of e-folds before the end of 
inflation must be greater than $50$ in most of the current models, and hence can hardly yield a sizable negative running following the relation \eqref{alpha_s}.

\section{Reheating in MBS}

Gravitational particle production in the MBS has been recently introduced in \cite{Quintin:2014oea}. The idea is the same as in inflationary models
with potentials without a minimum (the so-called non oscillatory models): to have an efficient reheating, one needs a non-adiabatic  transition in the 
expanding regime between two different phases in order to have enough gravitational particle creation. In inflation, there is an abrupt transition from a quasi de 
Sitter regime to a radiation-dominated one. During this transition, light particles are created and their energy density follows $\rho_r\sim a^{-4}$. On the other hand, after 
the end of the quasi de Sitter phase, the inflaton field, namely $\phi$, enters a kinetic-dominated period where the energy density of the inflaton field 
follows $\rho_{\phi}\sim a^{-6}$ \cite{Ford:1986sy,Peebles:1998qn}, which means that the inflation energy density decreases faster than that of radiation, and thus, the Universe 
becomes radiation dominated and matches with the hot 
Friedmann Universe. 

For the case of the MBS, the background equation is depicted by the improved Friedmann equation such as in holonomy corrected LQC. The non-adiabatic transition could happen in 
the contracting phase. In fact, a transition from matter-domination to an ekpyrotic phase with equation of state $P=\omega\rho$ where $\omega \gg 1$ could be assumed in the contracting 
regime. The resulting model is dubbed as the matter-ekpyrotic bounce \cite{Cai:2012va}. Since in the ekpyrotic phase the energy density of the field follows 
$\rho_{\varphi}\sim a^{-3(1+\omega)}$, which in the contracting phase increases faster than $a^{-6}$, anisotropies become negligible (note that the energy density of the 
anisotropies grows in the contracting phase as $a^{-6}$ which is faster than those of radiation and regular matter, and thus, without an ekpyrotic transition the isotropy of the 
bounce is destroyed, which is known as the BKL instability \cite{Belinsky:1970ew}). Moreover, the energy density of the field also increases faster than that 
of radiation, which means that the field dominates the evolution of the Universe in the contracting phase, but when the Universe bounces, radiation eventually dominates, because 
in the expanding phase $a^{-3(1+\omega)}$ decreases faster than $a^{-4}$. 

To be more specific, we will study reheating via massless $\chi$-particles nearly conformally coupled with gravity, using the method developed in \cite{Zeldovich:1971mw}. It is well
known that the energy density of the produced particles is related via the $\beta$-Bogoliubov coefficient as follows \cite{Ford:1986sy}
\begin{eqnarray}
 \rho_{\chi}=\frac{1}{2\pi^2 a^4}\int_{0}^{\infty}|\beta_{k}|^2k^3dk ~,
\end{eqnarray}
where
\begin{eqnarray}
 \beta_{k}=\frac{i(1-6\xi)}{2k}\int_{-\infty}^{\infty}e^{-2ik\eta}\frac{a''(\eta)}{a(\eta)}d\eta ~,
\end{eqnarray}
and where $\xi\cong \frac{1}{6}$ is the nonminimal coupling constant.

To perform the calculation we consider the simplest model of an abrupt transition from matter to ekpyrotic phase \cite{Cai:2012va},
\begin{eqnarray}
 a(t)=\left\{\begin{array}{ccc}
 a_E\left(\frac{t-t_0}{t_E-t_0}\right)^{2/3}& \mbox{fot} & t\leq t_E\\
 \left(\frac{3}{4}\rho_c(1+w)^2t^2+1\right)^{\frac{1}{3(1+w)}} &\mbox{for}& t\geq t_E ~,
 \end{array} \right.
\end{eqnarray}
where $w \gg 1$, $t_0=t_E-\frac{2}{3H_E}$,  $t_E$ is the time at which the transition occurs, and $H_E=H(t_E)$. Note that  $t_0$ has been chosen so that $a(t)$ has continuous 
first derivative at the transition time $t_E$.

First of all, to remove ultra-violet (UV) divergences one has to assume that the second derivatives of the scale factor are continuous at all times. In this case, after integration 
by parts the $\beta$-Bogoliubov coefficient becomes
\begin{eqnarray}
 \beta_{k} = -\frac{(1-6\xi)}{4k^2}\int_{-\infty}^{\infty}e^{-2ik\eta} \left(\frac{a''(\eta)}{a(\eta)}\right)'d\eta ~.
\end{eqnarray}
Now, for the sake of simplicity, we will assume as in \cite{Ford:1986sy} that the third derivative of the scale factor is discontinuous at the transition time $\eta_E$. Then, one 
has
\begin{eqnarray}\label{f1}
 \beta_{k} \cong \frac{9(1-6\xi)}{16ik^3}e^{-2ik\eta_E} w^2H_E^3a_E^3 ~.
\end{eqnarray}
As a consequence, the total energy density of the produced particles is
\begin{eqnarray}
 \rho_{\chi}\cong\frac{9}{16} (1-6\xi)^2 w^3\frac{\rho_E^2}{\rho_{pl}} ~,
\end{eqnarray}
where $\rho_E=3H_E^2$ is the energy density at the transition time.

In order to calculate the reheating temperature, namely $T_R$, first of all one has to define the reheating time $t_R$ as the time when the radiated energy 
density equals the background energy density. Since the background energy in the ekpyrotic phase for large values of $t$ is given by $\rho(t)=\frac{4}{3 w^2t^2}$, the 
reheating time is of the order
\begin{eqnarray}
 t_R\sim \sqrt{\frac{\rho_{ pl}}{\rho_E}}\frac{1}{w^{5/2}|1-6\xi|H_E} ~,
\end{eqnarray}
and thus, the reheating temperature is of the order \cite{Haro:2014wha}
\begin{eqnarray}
 T_R\sim \rho_{\chi}^{1/4}(t_R)\sim \sqrt{|1-6\xi|} w^{3/4}\lambda M_{pl} ~.
\end{eqnarray}
Note that we have written $\rho_E$ in terms of the Planck mass $M_{pl}\equiv\rho_{pl}^{1/4}$ as follows $\rho_E\equiv\lambda^2 M^4_{pl}$. Finally, choosing 
$\sqrt{|1-6\xi|} w^{3/4}\lambda\sim 10^{-7}$, this theoretical value can become consistent with the present observational bound \cite{Quintin:2014oea}.

\section{Challenges and problems of the MBS}

In the previous section, we have shown that the analysis in the plane $(n_s,\alpha_s)$ of the quasi-matter domination provides a range of values of the 
background field $\varphi$, whose corresponding theoretical values for the spectral index and its running fit well  the Planck2013 data. What is important to stress 
is that in the quasi-matter domination regime the value of the background field determines its derivative (\ref{x2}), which means that the analysis in 
the plane $(n_s,\alpha_s)$ provides a short curve in the phase space $(\varphi,\dot{\varphi})$ of initial conditions, whose corresponding 
orbits (solutions of the conservation equation) depict a Universe compatible with the data of the spectral index and its running provided by the 
Planck experiment. Consequently, for all these orbits one has to calculate (numerically) the amount of non-gaussianity and to compare it with the 
Planck data \cite{Ade:2013ydc,Ade:2015ava} in order to check which of these orbits are viable. This is a very complicated task 
as we will explain here.

\subsection{Non-gaussianities in the MBS}

In \cite{Cai:2009fn} the authors study non-gaussianities in the context of the MBS for a single scalar field using the formalism developed in \cite{Maldacena:2002vr}. In that study, 
the background solution depicts a matter dominated Universe in the whole contracting phase in the frame of GR. As usual, a nonsingular bounce may be achieved by including modified 
gravity effects or introducing nonconventional operators of the scalar field at high energy scales and hence it is expected that the nonlinear perturbations would only become 
important in the UV regime of bouncing cosmologies, which could be outside of the present observational window. For instance, a nonlinear treatment of primordial curvature 
perturbations passing through a nonsingular bouncing phase were numerically computed based on a particular model involving auxiliary ghost fields in \cite{Xue:2013bva}, of which 
the result reveal that 
there is no evidence pointing to a manifest generation of nonlinearities in the infrared regime. The difficulties of calculating the $3$-point correlation function during the 
bouncing phase were recently pointed out in \cite{Gao:2014hea} by studying a specific bouncing model: a background FLRW geometry with positive spatial curvature in the context 
of GR. One ought to be aware that this model differs from usual matter bouncing cosmologies since a quasi-matter domination in the contracting phase with positive spatial 
curvature does not give rise to a nearly scale invariant power spectrum.

At present, it remains an open issue to quantitatively calculate the amount of primordial non-gaussianities produced during the bouncing phase in a generic bounce model. 
Moreover, if one goes beyond GR with the flat FLRW geometry in order to achieve a bouncing background, for example using the holonomy corrected LQC or 
teleparallel $F(T)$ gravity, one has to use the very complicated second order perturbation equations in those theories, and try to 
calculate the $3$-point function for all the background orbits that are allowed by the $(n_s,\alpha_s)$ analysis.

%\vspace{0.5cm}

\subsection{Problems with particular bounce models}

Bounce models that depict the MBS with a single scalar field in LQC may also suffer from a gradient instability issue. For instance, 
dealing with holonomy corrected LQC (see \cite{Ashtekar:2011ni} for a recent review about non-perturbative LQC), the square of the sound speed 
is given by (see for instance \cite{Cailleteau:2011kr})
\begin{eqnarray}
 c_s^2=1-\frac{2\rho}{\rho_c} ~,
\end{eqnarray}
which means that in the bouncing phase (i.e., when the energy density satisfies $\rho_c/2<\rho\leq \rho_c)$, the square 
of the sound speed becomes negative. In spite of the fact that in holonomy corrected LQC, in order to calculate the power spectrum, only modes that satisfy 
the long-wavelength condition $|c_s^2k^2|\ll \left|\frac{z''}{z}\right|$ are used, it is important to realize that, in the super-inflationary regime, the 
UV modes (modes that satisfy the condition $|c_s^2k^2|\gg \left|\frac{z''}{z}\right|$) would suffer the Jeans instability. Moreover, in this super-inflationary 
regime, it is questionable the use of the linear perturbation equations, what could invalidate the results obtained about the value of the power spectrum because, as we 
have shown in formula (\ref{power}), it is calculated throughout the whole background regime. This issue is also associated with the unclear trans-Planckian physics that 
is expected to be described by the full theory of LQC \cite{WilsonEwing:2012bx, Rovelli:2013zaa}. However, one should be aware 
that these dangerous modes, since they are in the extremely UV scales, do not appear in the observable window of today's experiments.

Dealing with $F(T)$ and $F(R)$ gravity, the square of the sound speed is always positive. Thus, one can see that the issue of gradient instability 
is model dependent. However, as we have already explained in section $3$, teleparallelism suffers from the problem that the main invariant, the scalar 
torsion $T$ provided by the {\it Weitzenb\"ock} connection, is not at all an invariant, in the sense that it depends on the choice of the orthonormal 
basis in the tangent bundle (the tetrad). More precisely, a local Lorentz transformation in the tangent bundle applied to the orthonormal basis changes 
the value of the scalar torsion. In contrast, the scalar curvature $R$ provided by the {\it Levi-Civita} connection is a true invariant, but the problem with 
modified $F(R)$ gravity (recall that GR, i.e. $F(R)=R$, forbids bounces for geometries with a flat spatial curvature) is that the current very complicated bouncing 
models, obtained using the reconstruction technique \cite{Carloni:2010ph}, do not support a matter 
domination in the contracting phase, and thus the power spectrum is not scale invariant \cite{Bamba:2013fha}.

Another different possibility is to consider the MBS, for flat spatial geometries, in the context of GR by including extra matter fields, which can violate the Null Energy 
condition, such as a ghost condensate field \cite{Lin:2010pf} or a Galilean type one \cite{Qiu:2011cy, Easson:2011zy}. In these models, one ought to be aware
of the potential graceful exit problem as well as the gradient instability issue. For example, in a concrete cosmology of two-field matter bounce \cite{Cai:2013kja}, it was 
found that the model can be free from these dangerous issues with cosmological perturbations evolving through the nonsingular bouncing phase almost unchanged \cite{Battarra:2014tga}. 
Accordingly, the tensor-to-scalar ratio could be too large to explain the CMB observations and may require a curvaton mechanism \cite{Cai:2011zx} to give rise to an 
enhancement on curvature perturbations from entropy fluctuations.

%\vspace{0.5cm}

\section{Conclusions}

In this article we provided the main requirements that ought to be satisfied in a theoretical model of the MBS so that it could be a viable alternative 
to the inflationary paradigm. These requirements are from both observational and theoretical aspects. In particular, we performed a detailed analysis of the 
primordial perturbations in an extended scenario of the MBS, in which the equation of state parameter slightly deviates from zero in a dynamical way. Due 
to this dynamical deviation, we find that the model can naturally yield a non-vanishing spectral index $n_s$ and running parameter $\alpha_s$. As a result, the 
parameter space of the MBS models is greatly enriched. 

We applied this extended scenario into the theories of holonomy corrected LQC or teleparallel $F(T)$ gravity and then examined the values of tensor-to-scalar 
ratio $r$, as well as studied the gravitational reheating process. By combining the observational bounds upon the perturbation parameters $A_s$, $n_s$, $\alpha_s$ and $r$, we 
conclude that the present MBS models can be constrained very well. In particular, we explored the possibility of realizing a sizable negative running in bouncing cosmologies, which is 
expected to be tested in latest observational data (namely, the Planck data released in 2015 \cite{Planck:2015xua}). Additionally, we also commented on several unresolved 
issues including nonlinear perturbations during the bouncing phase,
the Jeans instability arising in the UV regime, and the Lorentz invariant in $F(T)$. These issues are not only important in constructing specific models of bouncing cosmologies, but 
they also put forward theoretical challenges to understanding quantum gravity theories.

\vspace{1cm}
{\bf Acknowledgments.} We thank Jaume Amor\'os, Sergei D. Odintsov, Jerome Quintin, and Edward Wilson-Ewing for their comments, which have been essential in order to improve 
this paper. The investigation of J. de Haro has been supported in part by MINECO (Spain), project MTM2011-27739-C04-01. YFC is supported in part by the Natural Sciences 
and Engineering Research Council (NSERC) of Canada and by the Department of Physics at McGill University.

\end{document}